# Oxynitride Thin Films versus Particle-Based Photoanodes: a Comparative Study for Photoelectrochemical Solar Water Splitting

Fatima Haydous,[†] Max Döbeli,[‡] Wenping Si,[†§] Friedrich Waag,[⊥] Fei Li,[†] Ekaterina Pomjakushina,[†] Alexander Wokaun,[∥] Bilal Gökce,[⊥] Daniele Pergolesi*[,†,∥] and Thomas Lippert*[,†,#,∇]

[†] Division for Research with Neutrons and Muons, Paul Scherrer Institut, 5232 Villigen-PSI, Switzerland

[‡] Ion Beam Physics, ETH Zürich, 8093 Zürich, Switzerland

[⊥] Center for Nanointegration Duisburg-Essen, Technical Chemistry I, University of Duisburg-Essen, 45141 Essen, Germany

[∥] Energy and Environment Research Division, Paul Scherrer Institut, 5232 Villigen PSI, Switzerland

[#] Laboratory of Inorganic Chemistry, Department of Chemistry and Applied Biosciences, ETH Zurich, 8093 Zürich, Switzerland

[∇] Molecular Photoconversion Devices Division, International Institute for Carbon-Neutral Energy Research (I2CNER) Kyushu University 744 Motooka, 819-0395 Fukuoka, Japan

**ABSTRACT:** The solar water splitting process assisted by semiconductor photocatalysts attracts growing research interests worldwide for the production of hydrogen as a clean and sustainable energy carrier. Due to their optical and electrical properties several oxynitride materials show great promise for the fabrication of efficient photocatalysts for solar water splitting. This study reports a comparative investigation of particle- and thin films-based photocatalysts using three different oxynitride materials. The absolute comparison of the photoelectrochemical activities favors the particle-based electrodes due to the better absorption properties and larger electrochemical surface area. However, thin films surpass the particle-based photoelectrodes due to their more suitable morphological features that improve the separation and mobility of the photo-generated charge carriers. Our analysis identifies what specific insights into the properties of materials can be achieved with the two complementary approaches.

## 1. INTRODUCTION

The possibility to use and store efficiently the virtually unlimited energy that we receive from the sun is the target of several research efforts since many decades.

One possible strategy is to convert solar energy into chemical energy by using visible light to split the water molecule and produce $O_2$ and $H_2$ gas. The latter can then be stored and used when and where is needed as a clean and renewable solar fuel. Even though this research field started back in 1972 with the pioneering work of Honda and Fujishima,[1] the interest in solar water splitting is still growing. This is mainly due to the necessity of clean and renewable energy resources to face the increasing energy demand, the depletion of fossil fuels and their unsustainable environmental impact.

Solar water splitting can be achieved by employing a semiconductor photocatalyst that is capable of absorbing the sunlight. When the semiconductor is irradiated, an electron-hole ($e^-$-$h^+$) pair is generated. The excited $e^-$ in the conduction band (CB) and the $h^+$ in the valence band (VB) contribute to the water reduction and oxidation reactions, respectively. Therefore, for efficient water splitting, the CB and VB of the semiconductor must be well positioned with respect to the water redox potentials.

Despite the chemical stability of oxide semiconductors and the suitable energy matching of the band edges with the water redox potentials, their performance remains restricted by their light absorption properties. In fact, due to the relatively large band gap, many oxide semiconductors can absorb photons mainly in the UV energy range which represents only a few percent of the solar spectrum. Exceptions are hematite ($\alpha$-$Fe_2O_3$) and $WO_3$ which have appeared to be promising visible-light absorbing photocatalysts. However, even though these materials have been extensively studied, the reported efficiencies are still too low to be used for practical applications. Both photocatalysts have CB edges lower than the $H^+/H_2$ redox potential which makes them only active for $O_2$ evolution.[2, 3] Thus, for overall water splitting, which involves the simultaneous evolution of stoichiometric amounts of $O_2$ and $H_2$, these photocatalysts must be coupled with other semiconductors that are active for $H_2$ evolution.

One approach for decreasing the band gap of oxides from the UV to the visible light energy range is through the complete or partial substitution of O by N to form (oxy)nitrides. The band gap reduction is mainly explained by the hybridization of O-2p and N-2p orbitals leading to an upward shift in the energy of the VB maximum. A recent work shows that the N substitution can also alter significantly the energy position of the CB minimum by affecting the lattice distortion and/or the electronegativity of the anions.[4] Oxynitrides with perovskite structure, such as $LaTiO_2N$,[5-7] $BaTaO_2N$,[8-10] $SrNbO_2N$[11-13] and $LaTaO_2N$[14, 15], having band gaps in the range between 1.8 and 2.6eV, have shown to be promising candidates for overall solar water splitting.

Oxynitride powders are typically fabricated by thermal treatment in ammonia of the precursor oxides. The photoactivity of samples based on particles is investigated either via photocatalytic or photoelectrochemical (PEC) measurements. In the former case, the amount of $O_2$ and $H_2$ evolved from the oxynitride powders are collected and the evolution rate is measured by gas chromatography.[16, 17] For PEC studies instead, oxynitrides are deposited on conducting substrates and used as the working electrode (the photoanode) of a three-electrode electrochemical cell filled with an aqueous electrolyte. Upon illumination, the holes oxidize water at the oxynitride surface to generate oxygen while the electrons move through the back contact to the counter electrode (the cathode, usually a Pt wire) where water is reduced to evolve hydrogen. The photocurrent measured between the cathode and the photoanode is proportional to the amount of evolved $O_2$ and $H_2$ gases.[6, 13, 14]

In most PEC investigations using oxynitride particles, the photoanodes are prepared by electrophoretic deposition (EPD).[6, 12, 14, 15, 18] This method is widely used for practical applications being relatively inexpensive and allowing the deposition of uniform coatings of particles over large areas. Additional post-deposition treatments are required to improve the overall electrical contact. A relatively large surface roughness, which enhances the electrochemical activity by widening the interface with the aqueous electrolyte, is easily achievable and mainly depends on the average particle size in the oxynitride powders. Other methods were also used for the preparation of particle-based photoanodes such as particle transfer method, which resulted in PEC performances higher than those of photoanodes prepared by EPD due to an enhanced electron transport in the photoanode.[5, 19]. Moreover, nanostructured oxynitride photoanodes are becoming more and more attractive as they outperform conventional oxynitride photoanodes prepared by EPD due to the increased surface area, improved electron-hole pair separation and enhanced electron transport.[8, 20]

In a different approach, the photoanode can be fabricated by thin film deposition technologies. By selecting the appropriate parameters for the thin film growth, fully-dense photoanodes can be fabricated with tunable crystalline and/or crystallographic properties. Polycrystalline and textured films with different average grain size can be used to probe the effect of crystallinity and crystalline defects on the PEC performance. The use of epitaxial thin films enables the fabrication of samples where different crystal planes, characterized by different charge transfer properties, face the electrolyte.[21, 22] The flat surfaces (nanometric surface roughness) of thin films can be used to design well-defined electrochemical active areas through controllable decoration with co-catalysts nanoparticles. These features make thin films ideal model systems to probe fundamental properties of materials not accessible using particle-based samples.

In spite of the undeniable interest, the literature dealing with the growth and characterization of oxynitride thin films for solar water splitting is still scarce. Oxynitride films were grown using different deposition methods such as chemical vapor deposition ($TiO_xN_y$[23]), atomic layer deposition (TiON[24]), sputtering ($LaTiO_xN_y$[22] and $TiO_xN_y$[25, 26]) and pulsed laser deposition ($TiO_xN_y$[27], $LaTiO_xN_y$[4, 21, 28, 29] and $BaTaO_2N$[30]). Among these studies, only three[21, 22, 29] report the investigation of the PEC performance of oxynitride thin films towards water splitting.

For this work, three perovskite oxynitrides, $LaTiO_xN_y$, $CaNbO_xN_y$ and $BaTaO_xN_y$, are used to prepare thin film and particle-based photoanodes for visible light-driven water splitting. EPD was used for the preparation of the photoanodes as this is the most frequently used method for the preparation of particle-based photoanodes. Herein we report a systematic comparison of the two sample designs for the three materials in terms of structural, optical and PEC properties. PEC measurements were chosen to investigate the photoactivity of the photoanodes. This is because for thin films the small roughness results in surface areas that are orders of magnitude smaller than for powders, thus making it difficult to measure the amount of $O_2$ and $H_2$ gases produced. Moreover, for the purpose of this study, the photoanodes are fabricated using the bare oxynitride semiconductors without any surface decoration/functionalization with co-catalysts. This makes the amount of evolved gases intrinsically small and difficult to detect. In a PEC measurement instead, photocurrents in the nA range and corresponding to very small amounts of evolved gasses can be accurately measured.

## 2. EXPERIMENTAL SECTION

**Preparation of oxynitride particle-based photoanodes.** The oxynitride powders were first synthesized by thermal ammonolysis of the oxides. All the oxide powders were prepared by solid state method. $Ca_2Nb_2O_7$ was prepared by the solid-state reaction between $CaCO_3$ and $Nb_2O_5$ in stoichiometric amounts at 1000 °C for 12 hours in the presence of a NaCl flux. $Ba_5Ta_4O_{15}$ was prepared by annealing a mixture of $BaCO_3$ and $Ta_2O_5$ at 1000 °C for 10 hours without using any flux. For the synthesis of $La_2Ti_2O_7$ powders, a stoichiometric mixture of $La_2O_3$ and $TiO_2$ was annealed at 1150°C in the presence of NaCl flux for 5 hours. In the case of $LaTiO_xN_y$ and $BaTaO_xN_y$, the oxynitride powders were obtained by ammonolysis at 950 °C for 11 and 10 hours, respectively, under an ammonia flow of 250 ml.min$^{-1}$. $Ca_2Nb_2O_7$ was annealed at 800 °C for 24 hours under ammonia flow (250 ml.min$^{-1}$) to produce $CaNbO_xN_y$.

The photoanodes were prepared via electrophoretic deposition. 40 mg of the oxynitride powder ($LaTiO_xN_y$, $CaNbO_xN_y$ or $BaTaO_xN_y$) were mixed with 10 mg of iodine in 50 ml of acetone followed by one hour sonication to obtain uniformly dispersed oxynitride powders. Electrophoretic deposition was conducted between two parallel fluorine-doped tin oxide (FTO) substrates (1×2 cm) placed in the oxynitride dispersion with a distance of 7 mm. The oxynitride powders were deposited on the negative electrode. The applied bias and deposition time were optimized for each oxynitride photoanode to result in the optimal PEC performance as shown in Figure S1 (Supporting Information). A bias of 20 V was applied for 3 min between the FTOs for the deposition of $LaTiO_xN_y$ and $CaNbO_xN_y$ photoanodes. In the case of $BaTaO_xN_y$, due to the smaller particle size, a bias of 30 V was required for 1 min. A post-necking treatment performed after the deposition of the oxynitride on the FTO substrates by dropping 30 μL of 10 mM $TaCl_5$ methanol solution on the photoanodes followed by drying in air. After repeating this cycle three times, the photoanodes were annealed at 300 °C for 30 min in air, followed by a heating cycle under ammonia flow for 1 hour at 450 °C.

**Preparation of oxynitride thin films.** The conventional PLD method was used in this work for the deposition of TiN buffer layers used as the current collector to enable the photoelectrochemical measurements of the oxynitride thin films. A commercially available TiN target was ablated. The deposition was done in vacuum at a base pressure of $10^{-6}$ mbar. The laser energy density was set between 3 and 3.5 $J.cm^{-2}$ with a laser repetition rate of 10 Hz. The films were deposited on (001)-oriented MgO substrate. The substrate-to-target distance was fixed at 50 mm. The thermal contact between substrates and heating stage was provided by Pt paste. The substrate temperature was set to 800 °C and was monitored by a pyrometer during the deposition.

For the deposition of the oxynitride films, the pulsed reactive-crossed beam laser ablation (PRCLA) method was used. Details about this technique can be found in ref.[31] Oxide targets were ablated in the presence of $NH_3$ pulsed gas jets to incorporate N into the films. Sintered $La_2Ti_2O_7$, $Ba_5Ta_4O_{15}$ and $Ca_2Nb_2O_7$ ceramic rod targets were prepared in our laboratory and used for the deposition of $LaTiO_xN_y$, $BaTaO_xN_y$ and $CaNbO_xN_y$ thin films, respectively. The deposition of oxynitrides was done in a $N_2$ background at a pressure of $8*10^{-4}$ mbar. $NH_3$ gas jets were injected near the ablation spot at the target. The delay time between the laser pulse and the gas jet was set to 30 µs. the duration of the gas jet was set to 400 µs. The laser was operated at a repetition rate of 10Hz. $LaTiO_xN_y$, $BaTaO_xN_y$ and $CaNbO_xN_y$ thin films were fabricated at a laser fluence of 3, 2.1 and 2.6 $J.cm^{-2}$, respectively.

A KrF excimer laser ($\lambda$=248 nm and pulse width=30 ns) was used for both, PLD and PRCLA. The substrates used for the deposition were ultrasonically cleaned with acetone and isopropanol.

**Characterization.** X-ray diffraction (XRD) was used for the structural characterization of powders and thin films. The measurements were conducted by a Bruker–Siemens D500 X-ray Diffractometer with a characteristic Cu $K_\alpha$ radiation. For the powders, $\Theta/2\Theta$ scans were performed to confirm the formation of a single phase of the oxides or oxynitrides. In the case of thin films, $\Theta/2\Theta$ scans were done in order to check the out-of-plane orientations of the oxynitride thin films. The crystalline quality was determined via the grazing incidence mode with $\Theta=1^o$.

The chemical composition and N content of the oxynitride thin films were measured using Rutherford backscattering (RBS) and Heavy-Ion Elastic Recoil Detection Analysis (ERDA). A 2 MeV $^4He$ beam was used for RBS measurements and a Si PIN diode was used as the detector. RUMP program was used to analyze the collected data. A 13 MeV $^{127}I$ beam was used for ERDA measurements with a gas ionization detector and a time-of-flight spectrometer.

For the oxynitride powders, the N content was obtained from thermogravimetric analysis. Aliquots of 20-60 mg of the oxynitride powders were heated in alumina crucibles to 1400 °C with a heating rate of 10 $°C.min^{-1}$ in 36.8 $mL.min^{-1}$ synthetic air. From the weight gain during the reoxidation of the oxynitride, the value of N content was calculated.

The absorbance of the particle-based photoanodes was determined from the measurement of the transmittance and reflectance. Diffuse reflectance was measured with a Cary 500 Scan UV-Vis-NIR spectrophotometer using an integrating sphere and the transmittance was collected with the same spectrophotometer in a dual beam configuration. For the thin films, the diffuse reflectance was not measured because of the smooth thin film surface and only transmittance measurements were acquired to calculate the absorbance. The films were deposited on double-side polished MgO substrates without the underneath TiN layer.

The morphologies of the thin films and particle-based photoanodes were investigated using a Zeiss Supra VP55 Scanning Electron Microscope. The SEM images were acquired using an in-lens detector with a typical acceleration voltage of 3 kV and a working distance of about 5 mm.

The thicknesses of the thin film samples were measured using a Veeco Dektak 8 stylus profilometer. The measurements were performed in contact mode with a scan speed of 1 mm/min using a stylus of tip diameter of 5 µm and a force of 8 mg between the tip and the sample.

BET measurements were conducted on a Quantachrome Nova 2200 at 77 K in the Center for Nanointegration Duisburg-Essen (CENIDE) at University of Duisburg-Essen to determine the surface area of the oxynitride powders.

PEC measurements were performed using a three electrode configuration with the oxynitride photoanode being the working electrode, a coiled Pt wire and Ag/AgCl (sat KCl) used as the counter and reference electrodes, respectively. The measurements were done in a quartz cell filled with 0.5 M NaOH electrolyte (pH=13). The electrolyte was purged with Ar for about one hour prior to the measurements. The oxynitride photoanodes were front-side irradiated with a 150 W Xe arc lamp (Newport 66477) equipped with AM 1.5G filter and with an output intensity of 100 $mW.cm^{-2}$ calibrated by a photodetector (Gentec-EO). The photocurrent measurements were acquired using a Solarton 1286 electrochemical interface.

## 3. RESULTS AND DISCUSSION

**Oxynitride particle-based photoanodes.** For the preparation of the particle-based photoanodes, $LaTiO_xN_y$, $BaTaO_xN_y$ and $CaNbO_xN_y$ powders were first prepared by thermal ammonolysis of $La_2Ti_2O_7$, $Ba_5Ta_4O_{15}$ and $Ca_2Nb_2O_7$ powders. The XRD patterns of the powders of the three oxynitrides and corresponding precursor oxides are presented in Figure 1. The XRD patterns match well with the expected crystal structures of the respective oxynitrides and oxides: the orthorhombic perovskite structures of $LaTiO_2N$ (ICSD Code: 168551) and $CaNbO_2N$ (ICSD Code: 55396), the cubic perovskite of $BaTaO_2N$ (ICSD Code: 202763), the monoclinic structures of $La_2Ti_2O_7$ (ICSD Code: 5416) and $Ca_2Nb_2O_7$ (ICSD Code: 26010), and the trigonal structure of $Ba_5Ta_4O_{15}$ (ICSD Code: 16028). In all XRD patterns, no secondary phases were observed.

The N content of the oxynitride powders was determined by thermogravimetric analysis from the mass difference between the oxynitride and the resulting oxide after heating in synthetic air. Figure S2 shows the results of these measurements. The thermogravimetric profiles agree with previous literature reports[32]: the oxynitride powders first gain weight due to O substitution in the anionic positions and partial N retention in form of interstitial molecules that bridges the metal cations. The weight reaches a peak before decreasing as the oxynitrides lose all N. For $LaTiO_xN_y$ and $CaNbO_xN_y$ the initial weight gain starts around 400 °C, while interestingly the final weight of the oxide is reached at much higher temperature for $LaTiO_xN_y$ than $CaNbO_xN_y$ (about 1200 °C and 700 °C, respectively). For both, the total mass gain is about 3.4% changing form the oxynitride to the oxide. A similar qualitative behavior was observed for $BaTaO_xN_y$ with an initial weight gain starting at about 600 °C and the final weight reached at 1200 °C. However, in this case, the total gain weight is only about 0.1%. This is because the mass change is directly proportional to the N content and inversely proportional to the molar mass of the resulting oxide. In the case of $BaTaO_2N$, the resulting oxide is $Ba_5Ta_4O_{15}$, whose molar mass is much higher than that of $Ca_2Nb_2O_7$ and $La_2Ti_2O_7$. The values of N content (y values) of $LaTiO_xN_y$, $BaTaO_xN_y$ and $CaNbO_xN_y$ estimated by thermogravimetric analysis are reported in Table 1.

**Table 1.** The properties of $LaTiO_xN_y$, $BaTaO_xN_y$ and $CaNbO_xN_y$ oxynitrides in terms of N content, band gap and BET surface area.

| Oxynitride | N content | Band gap [eV] | BET surface area [m2g-1] |
|---|---|---|---|
| $LaTiO_xN_y$ | 0.8 | 2.0 | 3.8 |
| $BaTaO_xN_y$ | 0.99 | 1.91 | 2.86 |
| $CaNbO_xN_y$ | 0.62 | 1.7 | 5.31 |

The absorbance of the oxynitrides and their corresponding oxides was determined from the diffuse reflectance measurements of the powders. As shown in Figure S3, $BaTaO_xN_y$ showed the most red-shifted absorption edge among the three oxynitrides. The band gap values of the oxynitrides, reported in Table 1, were determined from the analysis of the Tauc plots.

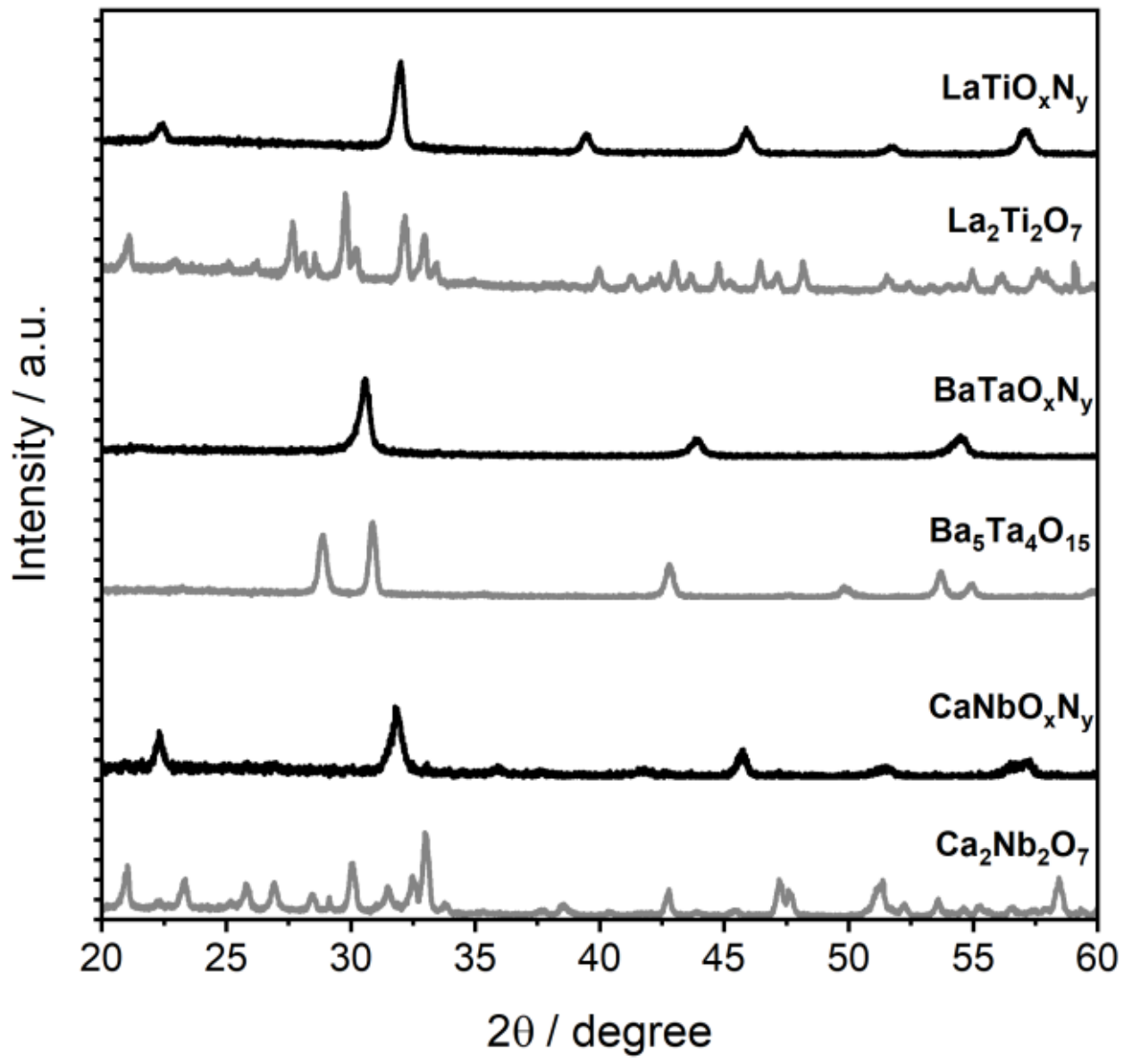

**Figure 1.** XRD patterns of $LaTiO_xN_y$, $BaTaO_xN_y$ and $CaNbO_xN_y$ oxynitride powders in black and their respective precursor oxides in grey**.**

As shown in Figure S3c and e, for $LaTiO_xN_y$ and $CaNbO_xN_y$, the band gap was determined from the direct transition to be 2.0 and 1.9 eV, respectively. For BaTaOxNy, the band gap was determined to be 1.7 eV for the indirect transition (Figure S3d).

Brunauer-Emmett-Teller (BET) measurements were used to determine the surface area of the oxynitride powders. The obtained surface area of the $LaTiO_xN_y$ oxynitride powders is about 3.5 times smaller than that reported in the literature for similarly prepared $LaTiO_xN_y$ powders.[33] For the $BaTaO_xN_y$ oxynitride prepared in this study from $Ba_5Ta_4O_{15}$, the measured BET surface area of 2.86 $m^2.g^{-1}$ is similar to that reported for $BaTaO_xN_y$ particles prepared from $Ba_2Ta_2O_7$ at the same ammonolysis temperature (~ 2 $m^2.g^{-1}$).[34] $CaNbO_xN_y$ powders prepared from a mixture of $CaCO_3$ and $NbCl_5$ at an ammonolysis temperature of 800 °C showed a surface area that is 6 times higher than that measured herein for $CaNbO_xN_y$.[17] The differences in BET surface area could originate from the different ammonolysis setups used. The resulting properties of the three oxynitrides are summarized in Table 1.

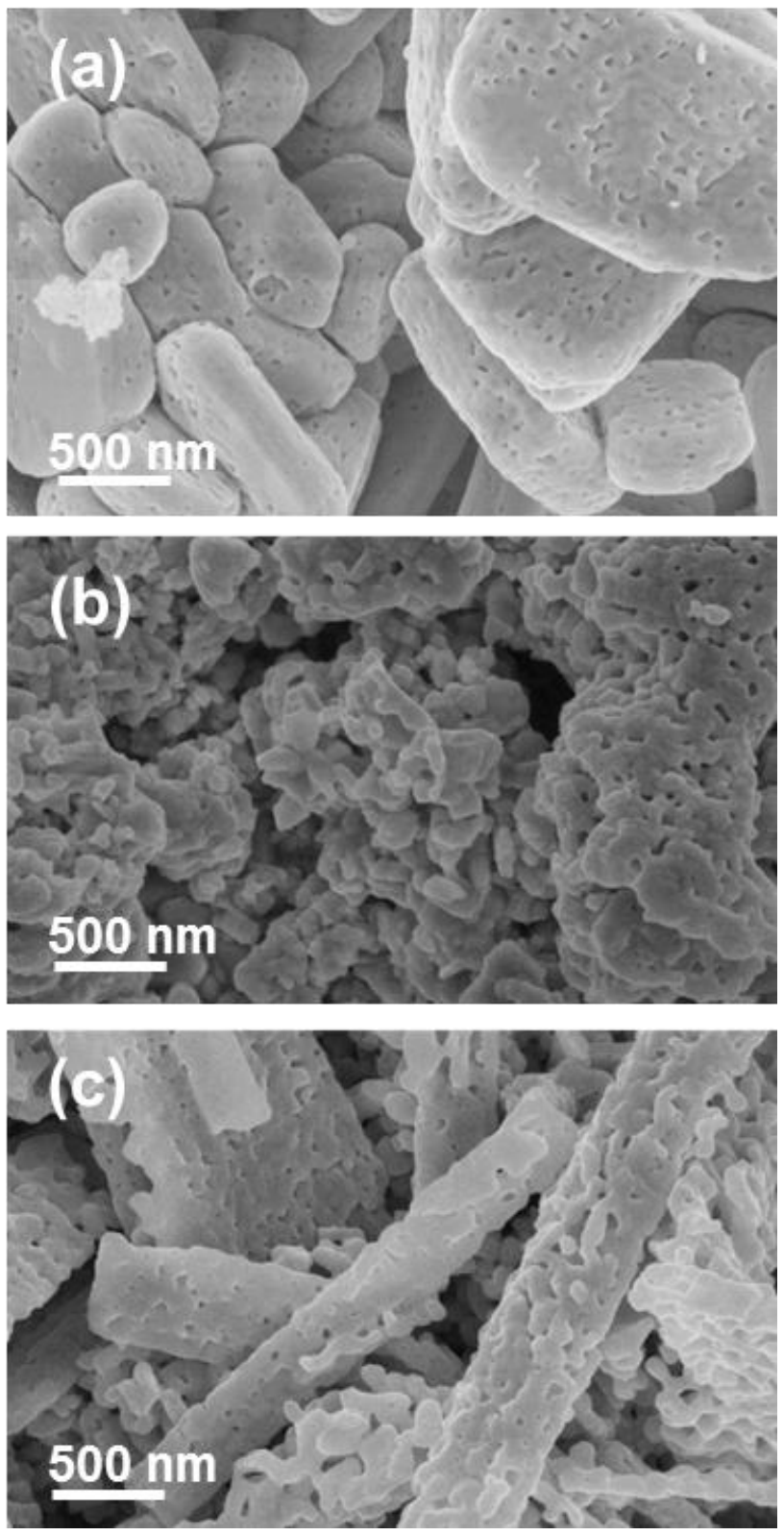

**Figure 2.** Top-view SEM images of (a) $LaTiO_xN_y$, (b) $BaTaO_xN_y$ and (c) $CaNbO_xN_y$ particle-based photoanodes.

The oxynitride photoanodes were then prepared by electrophoretic deposition on a fluorine-doped tin oxide (FTO) substrate. The deposition was followed by a post necking treatment to improve the electrical contact between the oxynitride particles as well as between the particles and the substrate. Briefly, a $TaCl_5$ film is deposited by drop-casting on the photoanodes. By annealing in air and $NH_3$, Ta(O,N) bridges are formed connecting the particles and enhancing the long-range mobility of the photogenerated charge carriers. This procedure is commonly used for the preparation of oxynitride photoanodes.[7, 10, 13]

Figure 2 shows the top-view scanning electron microscopy (SEM) micrographs of the obtained photoanodes. The $LaTiO_xN_y$ photoanode consists of porous particles with grain sizes in the range of 0.6 to 1.8µm. This is the typical morphology observed for $LaTiO_xN_y$ particles prepared by the ammonolysis of lanthanum titanium oxide produced via the solid-state route.[7] $BaTaO_xN_y$ appears as a dense agglomerate of much smaller particles with a diameter in the range of 50-100nm. Finally, the $CaNbO_xN_y$ photoanode consists of highly porous rectangular grains that are a few µm long and a few hundreds of nm wide. The porosity observed in these oxynitrides results from the substitution of $3O^{2-}$ by $2N^{3-}$ during the nitration reaction.[6]

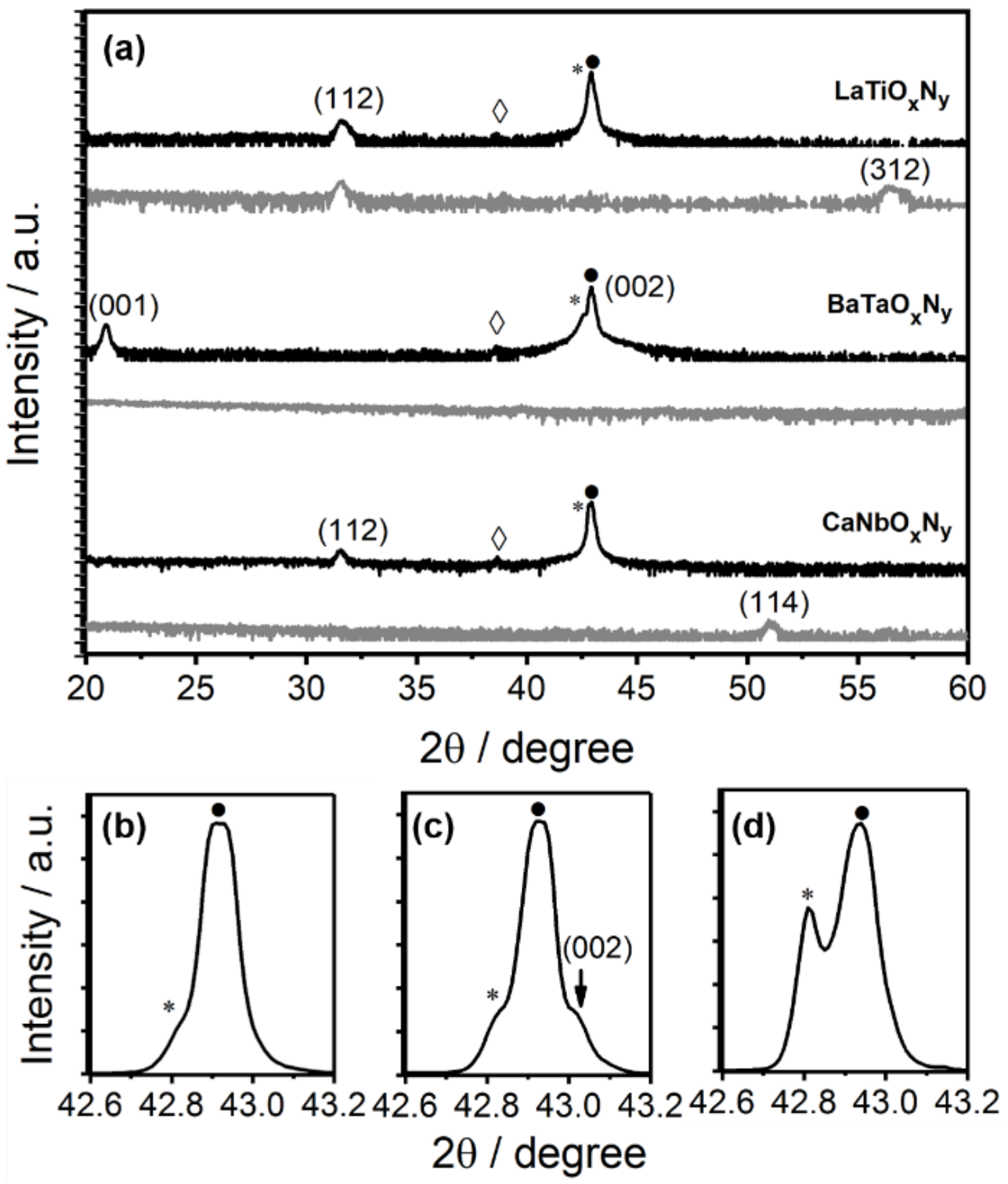

**Figure 3.** (a) XRD patterns of $LaTiO_xN_y$, $BaTaO_xN_y$ and $CaNbO_xN_y$ thin films. For each material, the black line shows the ϴ/2ϴ scan and the grey line shows the diffraction pattern acquired in grazing incidence mode. (b), (c) and (d) show an enlarged view of the ϴ/2ϴ scan plots in the angular range between 42.6o and 43.2o for $LaTiO_xN_y$, $BaTaO_xN_y$ and $CaNbO_xN_y$ thin films, respectively. The MgO substrate is marked by ●, TiN (002) diffraction peak is marked by *. In the ϴ/2ϴ scans the symbol ◊ indicates the $k_\beta$ reflex from the MgO substrate.

**Oxynitride thin film-based photoanodes.** Pulsed reactive crossed-beam laser ablation (PRCLA) was used to grow oxynitride thin films of $LaTiO_xN_y$, $BaTaO_xN_y$ and $CaNbO_xN_y$. The films were deposited on TiN-buffered single crystal MgO substrates (001)-oriented. The TiN buffer layer, deposited by conventional pulsed laser deposition (PLD) is used for this work as a current collector for the PEC measurements.[21, 29]

Figure 3a shows the XRD patterns for the three oxynitride thin films. Figure 3b-d show the enlarged view of the angular region around the (002) diffraction peak of the substrates. We notice first that the relatively small lattice mismatch between TiN and MgO drives the epitaxial growth of the buffer layers. The lattice parameters of MgO, TiN and the three oxynitrides are reported in Table2.

**Table 2.** Lattice Parameters of MgO substrate, TiN and the three oxynitrides: LaTiO2N, $BaTaO_2N$ and $CaNbO_2N$.

| Compound | Crystal Structure | Lattice Parameters [Å] | | |
|---|---|---|---|---|
| | | a | b | c |
| $LaTiO_2N$ | Orthorhombic | 5.603 | 5.571 | 7.880 |
| $BaTaO_2N$ | Cubic | 4.113 | - | - |
| $CaNbO_2N$ | Orthorhombic | 5.555 | 5.641 | 7.907 |
| MgO | fcc | 4.2113 | - | - |
| TiN | fcc | 4.2350 | - | - |

As can be seen in Figure 3, the $LaTiO_xN_y$ film shows only the (112) reflex in the ϴ/2ϴ scan. The same reflex is also observed in the grazing incidence mode together with the (312). These observations reveal the polycrystalline textured structure of the film. Similarly, a polycrystalline textured structure with preferential orientation along the (112) direction is obtained for the $CaNbO_xN_y$ film. We note that the (112) reflex is that with the largest relative intensity for these orthorhombic perovskite structures (ICSD Code: 168551 for $LaTiO_2N$ and ICSD Code: 55396 for $CaNbO_2N$).

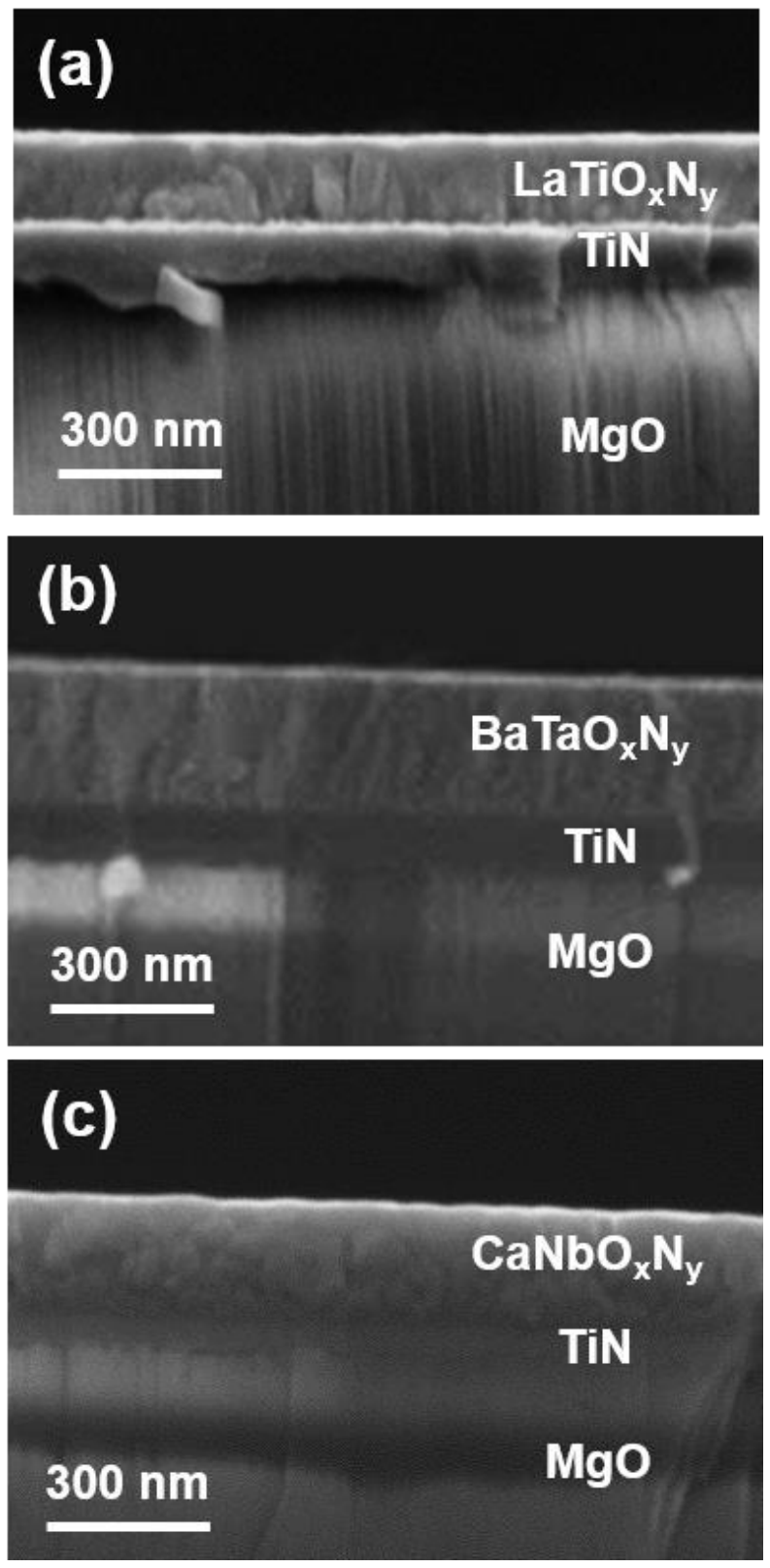

**Figure 4.** Cross-sectional SEM images of fractured (a) $LaTiO_xN_y$, (b) $BaTaO_xN_y$ and (c) $CaNbO_xN_y$ thin films grown on TiN-buffered MgO substrates.

In the case of $BaTaO_xN_y$, the films grow (001) epitaxially oriented on TiN/MgO substrates. In fact, the Θ/2Θ scan shows the (001) and (002) reflexes of the $BaTaO_xN_y$ cubic perovskite structure, whilst no diffraction peaks are visible in grazing incidence mode.

**Table 3.** Composition, N/O ratio and the thickness of the $LaTiO_xN_y$, $BaTaO_xN_y$ and $CaNbO_xN_y$ oxynitride thin films.

| Sample | Composition | Thickness [nm] | | |
|---|---|---|---|---|
| | | SEM | | Profilometry (TiN + Oxynitride) |
| | | TiN | Oxynitride | |
| $LaTiO_xN_y$/TiN/MgO | $La_{1.01}Ti_{0.99}O_{2.65}N_{0.34}$ | 120 | 152 | 270 |
| $BaTaO_xN_y$/TiN/MgO | $Ba_{1.07}Ta_{0.93}O_{2.25}N_{0.93}$ | 106 | 248 | 350 |
| $CaNbO_xN_y$/TiN/MgO | $Ca_{0.95}Nb_{1.05}O_{2.2}N_{0.65}$ | 70 | 215 | 280 |

Figure 4a, b and c show cross-section SEM images of $LaTiO_xN_y$, $BaTaO_xN_y$ and $CaNbO_xN_y$ thin films grown on TiN-buffered MgO. $LaTiO_xN_y$ and $CaNbO_xN_y$ films show the expected granular morphology of polycrystalline films. In the case of $BaTaO_xN_y$ instead, defined cleavage planes are visible with continuous fractures starting from MgO, passing through TiN and extending to $BaTaO_xN_y$ film. These are the typical morphological features of ordered epitaxial films.

Rutherford backscattering (RBS) was used to determine the cation and O content in the films with an experimental uncertainty of $\pm$3% for cations and $\pm$5% for oxygen. The nitrogen content of the films was calculated from the RBS results combined with Heavy-Ion Elastic Recoil Detection Analysis (ERDA) measurements from which the N to O ratio was obtained. The nitrogen content of the films was determined with an uncertainty of about $\pm$5%. The RBS spectra of the investigated oxynitride thin films on TiN-buffered MgO substrates are shown in Figure S4.

The (N+O) content of $LaTiO_xN_y$ is 2.99 which is nearly the ideal value of 3.0 for the perovskite phase. Also, the cation contents are close to the expected values. For the $CaNbO_xN_y$ films, the total anion content is 2.85. Even considering the uncertainty of the measurement, the anion deficiency seems to be significant. Moreover, RBS shows Ca deficiency and Nb enrichment. This can be explained by the much smaller atomic mass of Ca compared to Nb and the consequently different expansion profile of the Ca- and Nb-containing species in the plasma plume. The deficiency of the lighter elements in the chemical content of the film is often associated with the PLD process and was observed for instance in $La_{1-x}Ca_xMnO_3$ and $La_{1-x}Ca_xCoO_3$ films where the films were enriched with La and deficient in Ca.[35, 36] For the $BaTaO_xN_y$ films, a (N+O) content of more than 3.0, along with a Ba enrichment at the expense of Ta is observed. A Ba-enriched oxide was used for the synthesis of the powders as well as for the fabrication of the target for PLD due to the high volatility of Ba. The RBS measurements of our films show that further optimization of the deposition conditions and/or the Ba content of the target material is required to achieve the desired 1:1 ratio of the cation content. Overall, however, the XRD analysis in Figure 3 shows that for the three materials under investigation the measured deviation of chemical compositions has no influence on the crystalline structures, neither lead to the formation of secondary phases.

For the thin films of the three oxynitrides, Table 3 reports the chemical composition estimated by RBS and ERDA and the thickness measured by SEM. The total thickness of the oxynitride film and the underlying TiN buffer layer, as determined by profilometry, is in good agreement with the SEM measurements.

**Photoelectrochemical characterization.** In order to compare the photoactivity of oxynitride photoanodes based on thin films and particles, photoelectrochemical (PEC) measurements were performed by conducting current-voltage scans in three-electrode configuration. For the three materials and for both particle- and film-based photoanodes, the electronic current decreased to 50 – 30% of its initial value during the first potentiodynamic scans. After a few (three to six) potentiodynamic tests the current reaches stable and reproducible values. As reported in previous studies, this is largely due to the initial oxidation of the surface of the samples driven by the electrochemical process.[21] In the selected experimental setup, the initial surface oxidation can, in fact, generate a current density of tens of $\mu A.cm^{-2}$ which is in the same range of the photocurrent density typically measured using the bare oxynitride semiconductor (without co-catalyst decoration). For a meaningful comparison of the actual photoactivity of the samples, hereinafter we report the values of the stabilized current which is measured acquiring potentiostatic scans at different potentials. In the potentiostatic scans the voltage is held constant for 300 s before acquiring the value of the corresponding current. These are the values of current that we ascribe to the actual photoactivities of the samples and that we use here to compare the PEC activities of the three materials and the two sample designs (particle and film). The photocurrent densities are then calculated by subtracting the value of the current obtained without illumination (dark current) and normalizing the photocurrent values to the illuminated area of the sample. The dark current densities of the three oxynitride thin film and particle-based photoanodes are shown in Figure S5. Figure 5 shows the photocurrent density values obtained for $LaTiO_xN_y$, $BaTaO_xN_y$ and $CaNbO_xN_y$ particle-based (a) and thin film (b) photoanodes at potentials between 1.1 and 1.5 V vs RHE.

The literature on PEC characterization of oxynitride thin films is basically restricted to our previous study[21] and the study reported by Le Paven-Thivet and co-workers[22], both on $LaTiO_xN_y$ films. These two studies are in agreement with the results presented here. Concerning $BaTaO_xN_y$ and $CaNbO_xN_y$ instead, to our knowledge the PEC characterizations in Figure 5b for the thin films are the first reported in the literature.

Concerning the PEC characterizations of the particle-based photoanodes, the literature typically reports potentiodynamic measurements of devices optimized for maximum performance by thermal treatment in $H_2$, surface passivation with additional coatings and decoration with co-catalyst nanoparticles acting as the electrochemically active sites. [7, 9, 37, 38] Such treatments can improve the photoactivity by orders of magnitudes and in some cases these modifications of the surface are essential to enable the functionality of the material which is otherwise inactive or unstable. However, it has to be considered that the effects of the aforementioned treatments are different for different materials and could be different for particle- and film-based photoanodes. This would make more difficult the direct comparison of the two sample designs to understand if and how the film-based electrodes can be used as investigation tools complementary to particle samples. For these reasons, in the present study, the bare semiconducting oxynitrides are used.

Figure 5a shows the stabilized photocurrent densities measured for the three oxynitride particle-based photoanodes. At 1.5 V vs. RHE, for example, the current densities were about 30, 10 and 5 $\mu A.cm^{-2}$ for $LaTiO_xN_y$, $BaTaO_xN_y$ and $CaNbO_xN_y$ respectively. At 1.2 V vs. RHE the current densities decreased to about 10, 5 and 3 $\mu A.cm^{-2}$.

Using the bare $LaTiO_xN_y$ semiconductor, Landsmann and coworkers[7] reported values of photocurrent densities ranging from 20 to 120 $\mu A.cm^{-2}$ at 1.2 V vs. RHE depending on the fabrication method of the powders that led to BET surface area ranging from about 8 to 14 $m^2.g^{-1}$. At the same voltage, our samples show a photocurrent density of about 10 $\mu A.cm^{-2}$. Considering that this is the result of potentiostatic characterizations (instead of potentiodynamic) and taking into account the smaller BET surface area of about 4 $m^2.g^{-1}$ measured for our $LaTiO_xN_y$ samples, we conclude that the photocurrent density measured for our $LaTiO_xN_y$ samples is in line with literature reports.

Regarding $BaTaO_xN_y$ powders, the PEC characterization of the bare semiconductor is not reported in the literature. Using $BaTaO_xN_y$ particles, Higashi and colleagues[9] studied the combined effects of thermal treatments and co-catalysts. The optimized sample allowed achieving the very high value of photocurrent density of about 3.5 $mA.cm^{-2}$ at 1.2 V vs. RHE. In that study, however, current density values lower than about 100 $\mu A.cm^{-2}$, as those attainable using the base oxynitride, were not discussed.

Finally, concerning $CaNbO_xN_y$ to our knowledge the present study reports the first PEC characterizations of photoanodes based on this oxynitride. Computational studies have showed that this oxynitride can be used for both water oxidation and reduction reactions.[39] In addition, oxygen and hydrogen gases were experimentally shown to evolve from $CaNbO_xN_y$ particles suspended in aqueous solution.[17] It was shown that this material is active for both $O_2$ and $H_2$ evolution from aqueous $AgNO_3$ and methanol, yet it is important to control its preparation procedure because $NbO_xN_y$ can be formed easily as a byproduct that is detrimental to its photoactivity.[17] For the $CaNbO_xN_y$ samples prepared for this work, the XRD analysis shows no evidence of formation of secondary phases.

Among the three materials, $CaNbO_xN_y$ exhibited the lowest performance for both thin film and particle-based photoanode (Figure 5). This could be due to the higher density of defects present in this material compared to the other two oxynitrides. This hypothesis is supported by the higher background absorbance measured for the photoanodes composed of this oxynitride in comparison to $LaTiO_xN_y$ and $BaTaO_xN_y$ as can be seen in Figure 6.

Regarding the thin film samples of the three materials, the comparison of Figure 5a and b clearly shows that the values of photocurrent densities measured with the thin films lay in a range at least one order of magnitude lower than that for particles. This is due to the fact that the photocurrent density is calculated considering the illuminated area of the sample under investigation. While for thin films the illuminated area and the electrochemical area coincide, for particle-based samples the latter can be much wider due to the higher roughness. Taking into account a loading density of 0.4-0.45 mg.cm$^{-2}$ that is typical for powder samples,[9, 18] and using the BET surface area of the oxynitride powders reported in Table 1, the electrochemical area of particle-based photoelectrodes is estimated to be 15-17, 12-14 and 20-23 times larger than that of thin films for $LaTiO_xN_y$, $BaTaO_xN_y$ and $CaNbO_xN_y$, respectively. The higher surface area per illuminated area of the particle-based photoanodes results in larger number of active sites for hole injection to the electrolyte and thus more sites for oxygen evolution. This geometrical factor explains the one order of magnitude higher absolute performance of particle-based photoanodes compared to films.

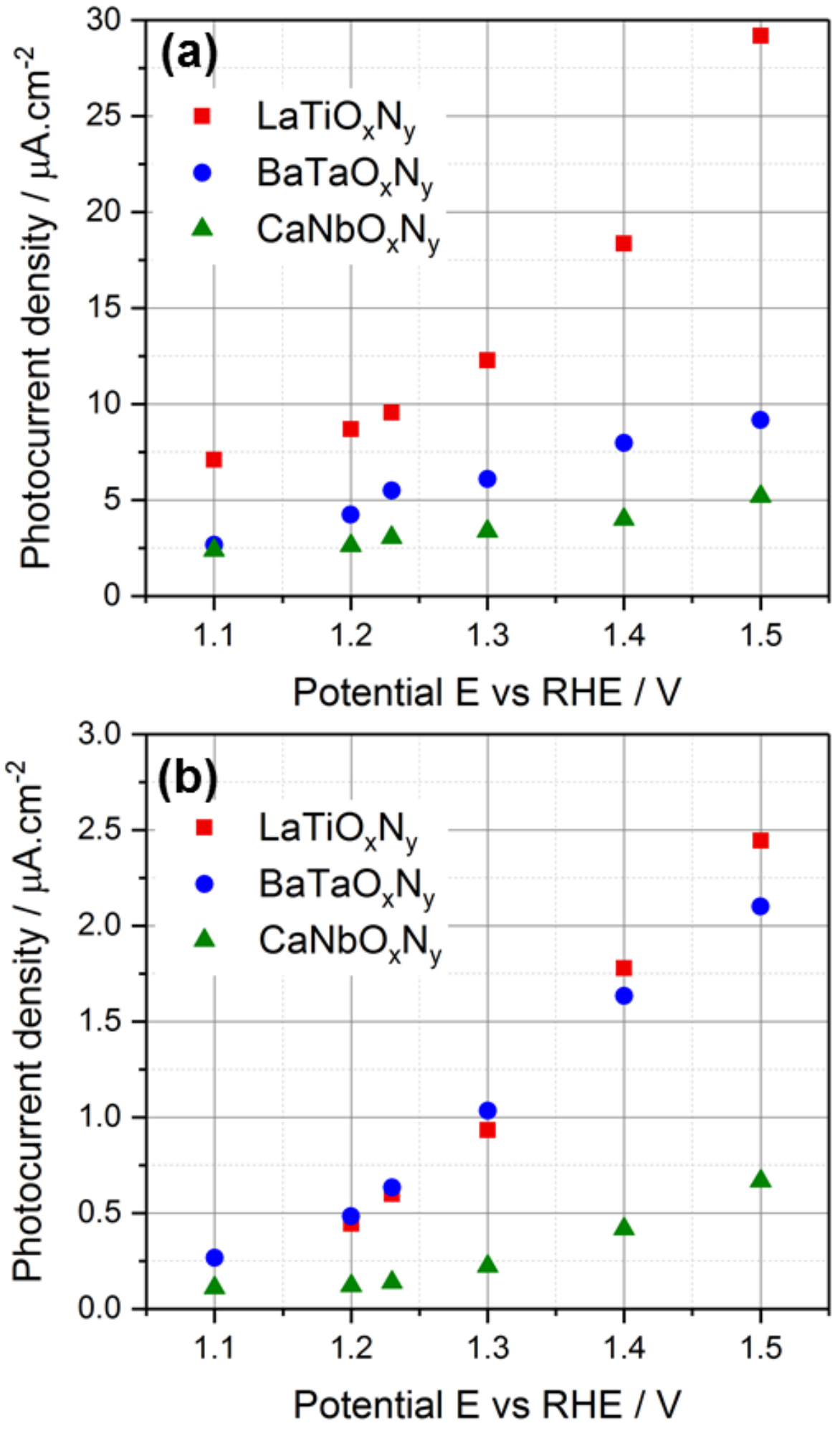

**Figure 5.** Photoelectrochemical performance of oxynitride (a) particle-based photoanodes and (b) thin films. The red squares correspond to $LaTiO_xN_y$ photoanodes, blue circles to $BaTaO_xN_y$ and green triangles to $CaNbO_xN_y$.

Another factor which explains the higher photocurrent achieved by the particle-based photoanodes compared to thin films is the difference in absorbance. The absorbance spectra of the oxynitride photoanodes are presented in Figure 6. The absorbance of the particle-based photoanodes was calculated from the reflectance and transmittance measurements. For the thin films, the diffuse reflectance was not considered due to the smoothness of the surfaces. For the transmittance measurements, oxynitride thin films were deposited on double-side polished MgO substrates without a TiN buffer layer underneath. All other deposition parameters were the same as for the films used for the PEC characterizations.

Figure 6 shows that for the three oxynitrides, particle-based photoanodes have higher absorption in the whole UV-Vis range compared to thin films. As mentioned, $CaNbO_xN_y$ photoanodes show high background absorption, possibly due to the higher defect density leading to higher recombination rate. Further investigation are planned to verify this hypothesis and understand whether it is possible to improve the performance of this material.

The higher light absorption of particles can be due to compositional or morphological reasons. For example, by comparing Tables 1 and 3 it can be observed that, while for $CaNbO_xN_y$ and $BaTaO_xN_y$ the N content of particles and thin films is very similar, for $LaTiO_xN_y$ the thin films have significantly lower N content than the powders. A lower than stoichiometric N content widens the band gap, thus shifting the photoresponse toward the UV energy range, i.e. the $LaTiO_xN_y$ thin films absorb fewer photons than the particle-based samples in the visible. On the other hand, even when the chemical composition is very similar, as in the case of $BaTaO_xN_y$ films and particles, the thickness of the oxynitride layer can affect the light absorption properties. All particle-based photoanodes are a few µm thick compared to 150-250 nm that is the thickness for the thin films. Therefore, it is expected that more light is absorbed with the particle-based samples even though this does not necessarily lead to better performance since the recombination rate play a crucial role. The effect of thickness is clearly measurable not only by comparing thin films and particle-based photoanodes but also particle-based photoanodes with different thicknesses. As an example, $CaNbO_xN_y$ particle-based photoanodes prepared under different electrophoretic deposition times with thicknesses varying between 0.5µm to 4.9µm showed different absorbance. As presented in Figure S6, the transmittance of the photoanodes decreased with increasing the thickness.

We conclude that the morphological differences in terms of thickness (that affect the light absorption properties) and surface roughness (that determine the BET area) can account for the lower absolute value of photocurrent densities measured with the thin film samples. It is worth noticing that, by considering the BET factor in the calculation of the current density, the values obtained for the thin films are nearly the same, or even higher, than those calculated for the particle-based samples (Table S1).

It is worth noticing that $BaTaO_xN_y$ thin films showed similar performance to $LaTiO_xN_y$ thin films (Figure 5b), while $BaTaO_xN_y$ particles showed photocurrent densities at least two-times smaller than $LaTiO_xN_y$ over the all range of applied potentials (Figure 5a). The comparison of Figure 2 and Figure 4 shows that morphologically the two thin films are indeed very similar, while the two particle samples looked very different, $LaTiO_xN_y$ showing a much larger average grain size. The larger average grain size of $LaTiO_xN_y$ appears to be highly beneficial. The use of thin film samples allows the comparison of PEC properties of materials with the same morphology, thus independently on the specific synthesis properties.

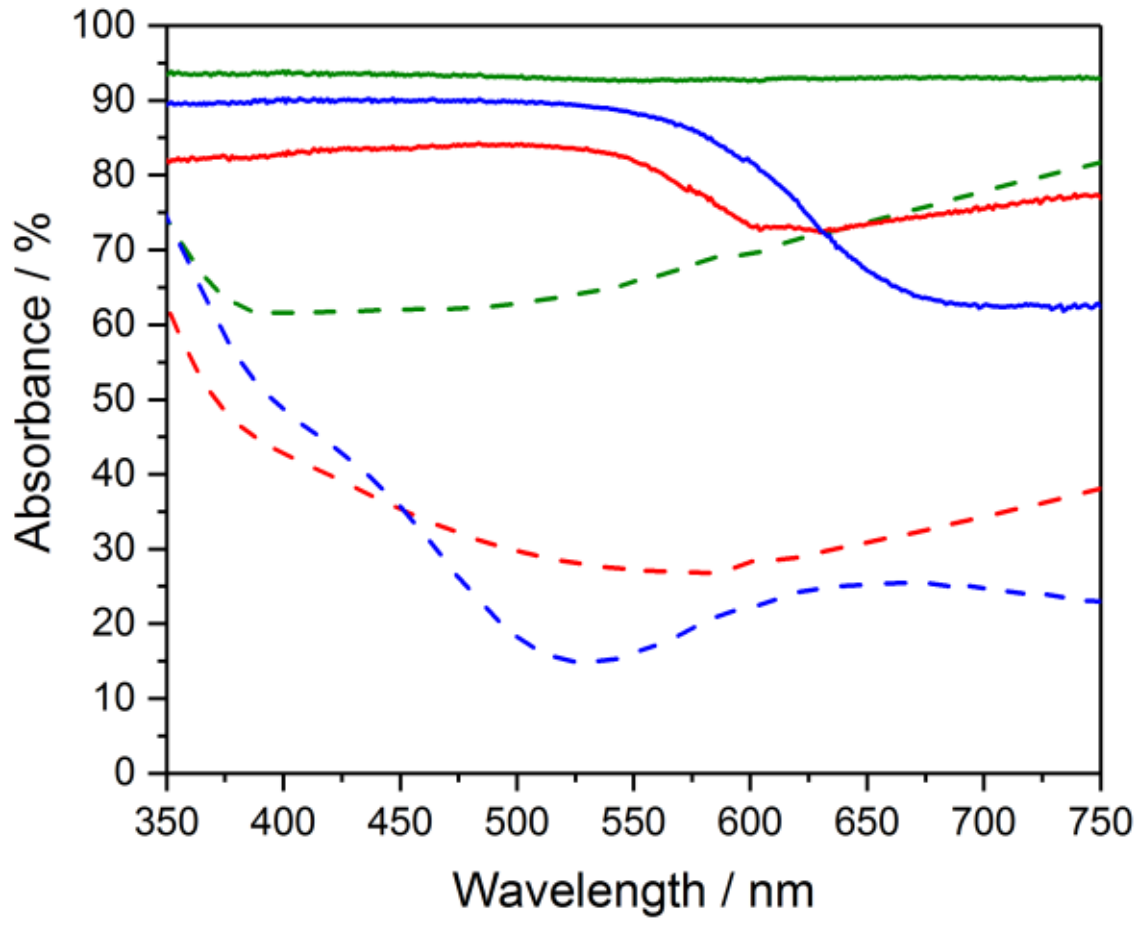

**Figure 6.** Absorbance spectra of $LaTiO_xN_y$ (in red), $BaTaO_xN_y$ (in blue) and $CaNbO_xN_y$ (in green) photoanodes. The solid curves correspond to the absorbance of the particle-based photoanodes and the dotted curves are for the thin films.

The schematics of the two photoanode systems, thin film and particles, are shown in Figure 7. Thin films consist of relatively large grains in good contact with each other. Particle-based photoanodes instead are composed of agglomerates of powders with smaller grains electrically not well connected and a post necking is needed to bridge them and enhance the mobility of the photogenerated carriers.

The much higher surface roughness of the particle-based samples is crucial to increase performance, not to mention the comparably inexpensive fabrication method. However, thin film technologies allow the growth of samples with well-defined, atomically flat surface, controlled crystallographic property and fully dense bulk morphology. Thin films are ideal tools to achieve surface sensitivity for the characterization of the electrode-electrolyte interface. The flat surface may allow the controlled decoration with co-catalyst nanoparticles. The epitaxial growth allows probing the effect of different crystallographic surface orientations. The fully dense morphology makes the PEC test independent on the synthesis process.

These are some of the most relevant applications of thin films as complementary investigation tools for the development of oxynitride photoanodes for solar water splitting.

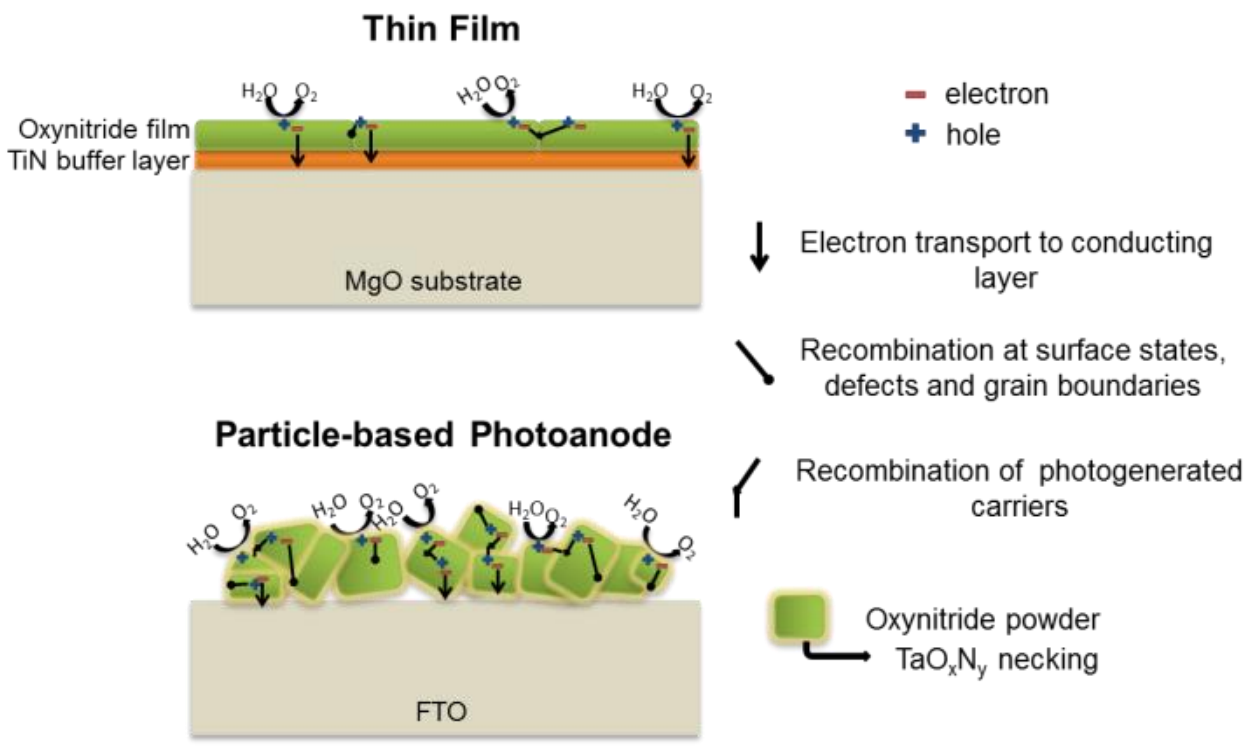

**Figure 7.** Schematics of thin films and particle-based photoanodes.

## 4. CONCLUSION

In this manuscript we compare the photoelectrochemical properties towards visible light-driven water splitting of different oxynitride photoanodes based on thin films and particles. This study aims at understanding to what extent the thin film systems can be used to complement the standard investigation approach.

We propose the use of potentiostatic photoelectrochemical characterizations, as opposed to the typically used potentiodynamic measurements, for a meaningful comparison of the photoelectrochemical activities toward solar water splitting of the bare semiconductor materials.

A rational comparison of the two sample designs reveals that the intrinsic morphological differences (surface roughness and thickness) can be accounted for by normalizing the measured photocurrent to the effective surface area and considering the difference in absorbance.

The results show that for the three oxynitrides, the particle-based samples achieved the higher absolute values of photocurrent densities, mainly due to the much larger surface area. Nevertheless, the higher photocurrent densities measured with the particle-based samples can be accounted for by considering the BET surface area and the different absorbance. Indeed, the better crystalline quality and the good electrical contact between grains make the thin films morphology much better suited to facilitate separation and migration of the charge carriers.

This study highlights the usefulness of coupling and properly comparing the results obtained with these two complementary approaches. Once understood how the thin films relate to the respective particles, the thin films with their well-defined surfaces become ideal experimental platforms for surface studies.

The advantages of both sample designs might be coupled by growing crystalline and fully dense films on 3D nanostructures to widen the electrochemical area of the films to values similar to particles or even larger still preserving their bulk properties. This is the topic of future investigations.

## ASSOCIATED CONTENT

### Supporting Information

Potentiodynamic scans of $LaTiO_xN_y$, $BaTaO_xN_y$ and $CaNbO_xN_y$ particle-based photoanodes under varied deposition conditions, thermogravimetric curves of the three oxynitride powders, the absorbance spectra of the oxides and oxynitrides and the Tauc plots for the latter, the RBS spectra of the $LaTiO_xN_y$, $BaTaO_xN_y$ and $CaNbO_xN_y$ oxynitride thin films, the dark current densities of the oxynitride particle-based and thin film photoanodes, transmittance spectra of the $CaNbO_xN_y$ particle-based photoanodes with different thicknesses and table showing the photocurrent densities of the photoanodes normalized to their BET factors. This material is available free of charge via the Internet at http://pubs.acs.org.

## AUTHOR INFORMATION

### Corresponding Author

* Emails: daniele.pergolesi@psi.ch, thomas.lippert@psi.ch

### Present Addresses

§ (W. Si) Key Laboratory of Advanced Ceramics and Machining Technology, Ministry of Education, School of Materials Science and Engineering, Tianjin University, Tianjin 300072, P.R. China.

## ACKNOWLEDGMENT

The authors would like to thank Paul Scherrer Institut, the NCCR MARVEL, funded by the Swiss National Science Foundation, and the Swiss Excellence Governmental Scholarship for the financial support of this work. Fatima Haydous was a Swiss Government Excellence Scholarship holder for the academic years 2014-2017 (ESKAS No. 2014.0282).

Table of Contents

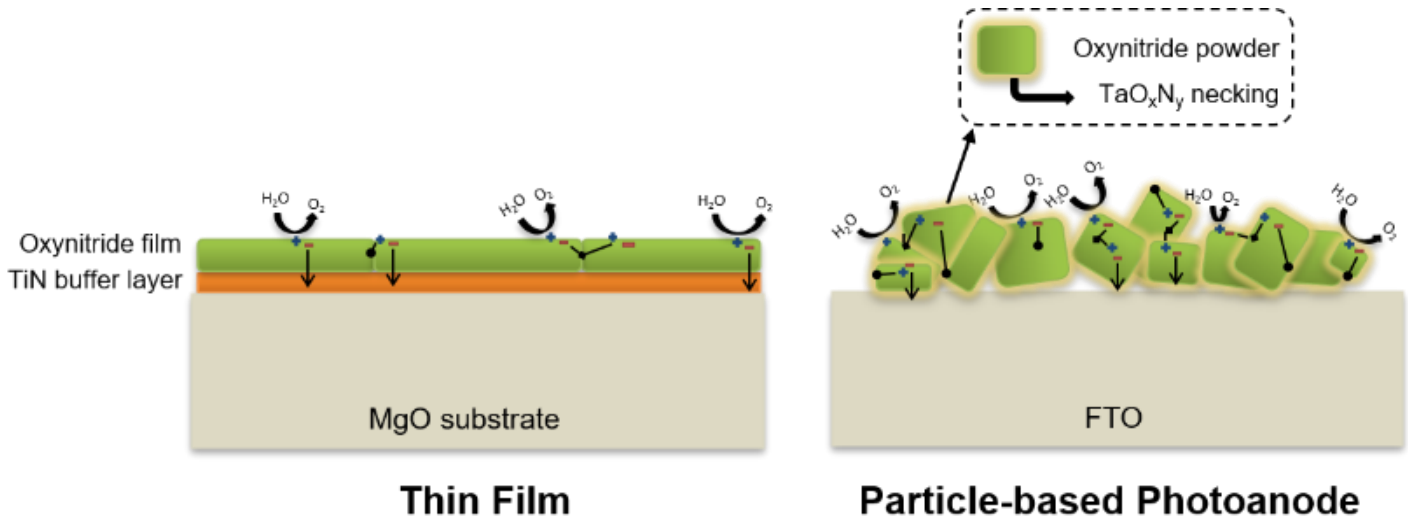

# Supporting Information

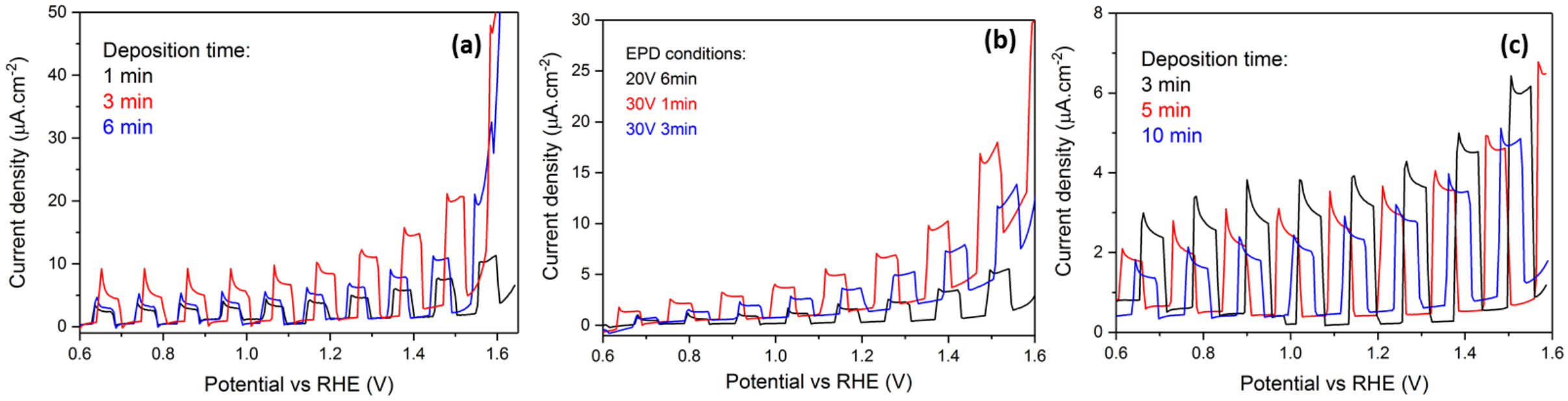

Figure S1. Third potentiodynamic scans acquired for (a) $LaTiO_xN_y$, (b) $BaTaO_xN_y$ and (c) $CaNbO_xN_y$ particle-based oxynitride photoanodes deposited by EPD under different conditions. The values of photocurrent densities above 1.5 V vs. RHE may be affected by large uncertainty due to the increasingly large contribution of the dark-current.

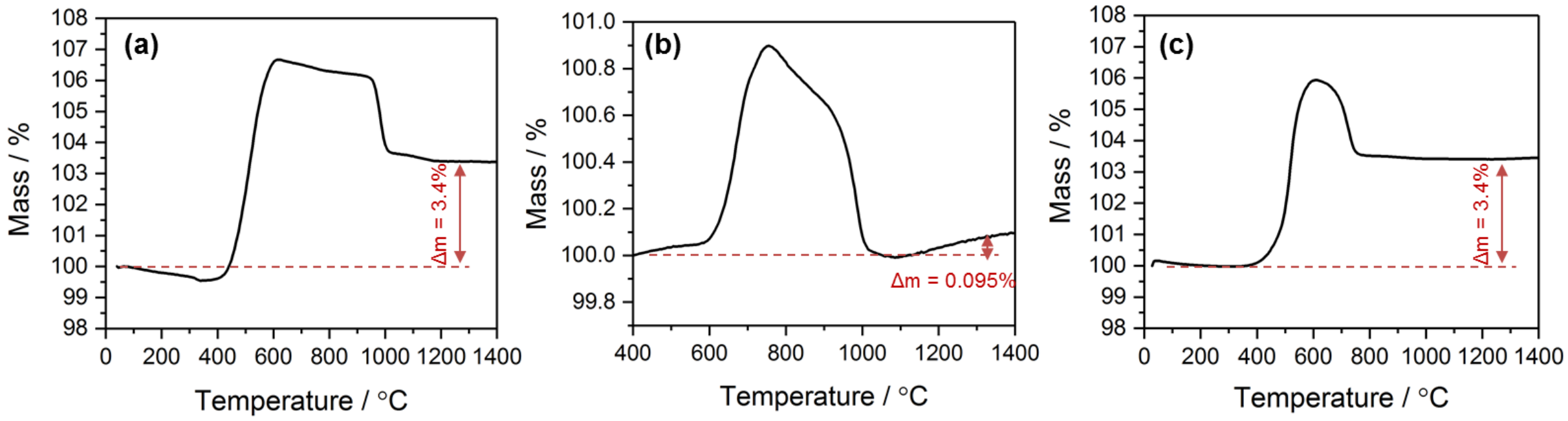

Figure S2. Thermogravimetric (TG) curves of (a) $LaTiO_xN_y$, (b) $BaTaO_xN_y$ and (c) $CaNbO_xN_y$ oxynitride powders.

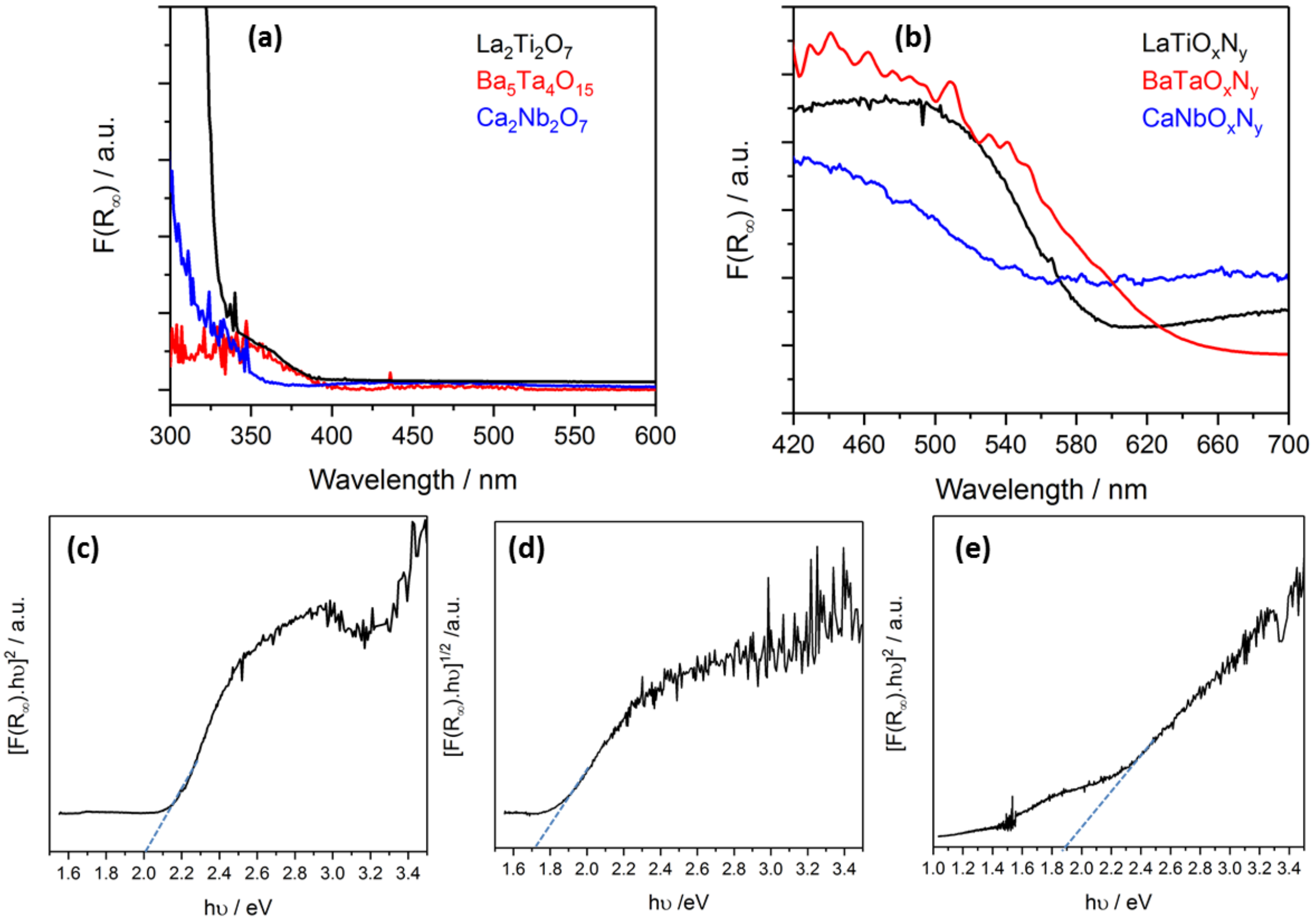

Figure S3. The absorbance spectra of (a) the oxides: $La_2Ti_2O_7$, $Ba_5Ta_4O_{15}$ and $Ca_2Nb_2O_7$ and (b) their oxynitrides: $LaTiO_xN_y$, $BaTaO_xN_y$ and $CaNbO_xN_y$. (c), (d) and (e) are the Tauc plots of $LaTiO_xN_y$, $BaTaO_xN_y$ and $CaNbO_xN_y$, respectively.

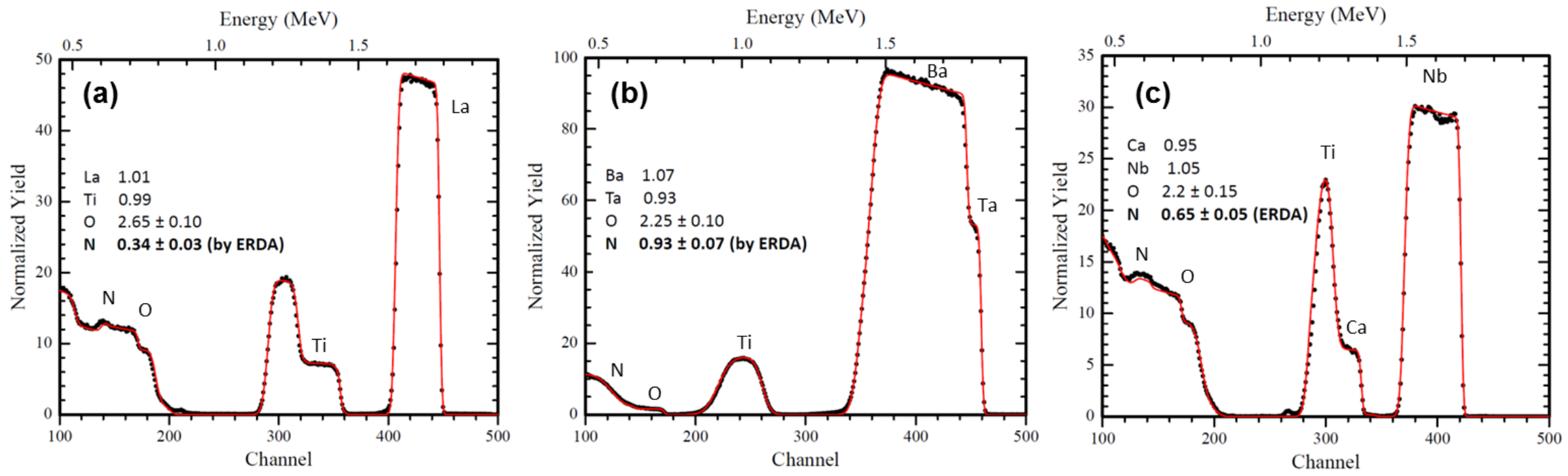

Figure S4. RBS spectra of (a) $LaTiO_xN_y$, (b) $BaTaO_xN_y$ and (c) $CaNbO_xN_y$ thin films grown on TiN/MgO substrate.

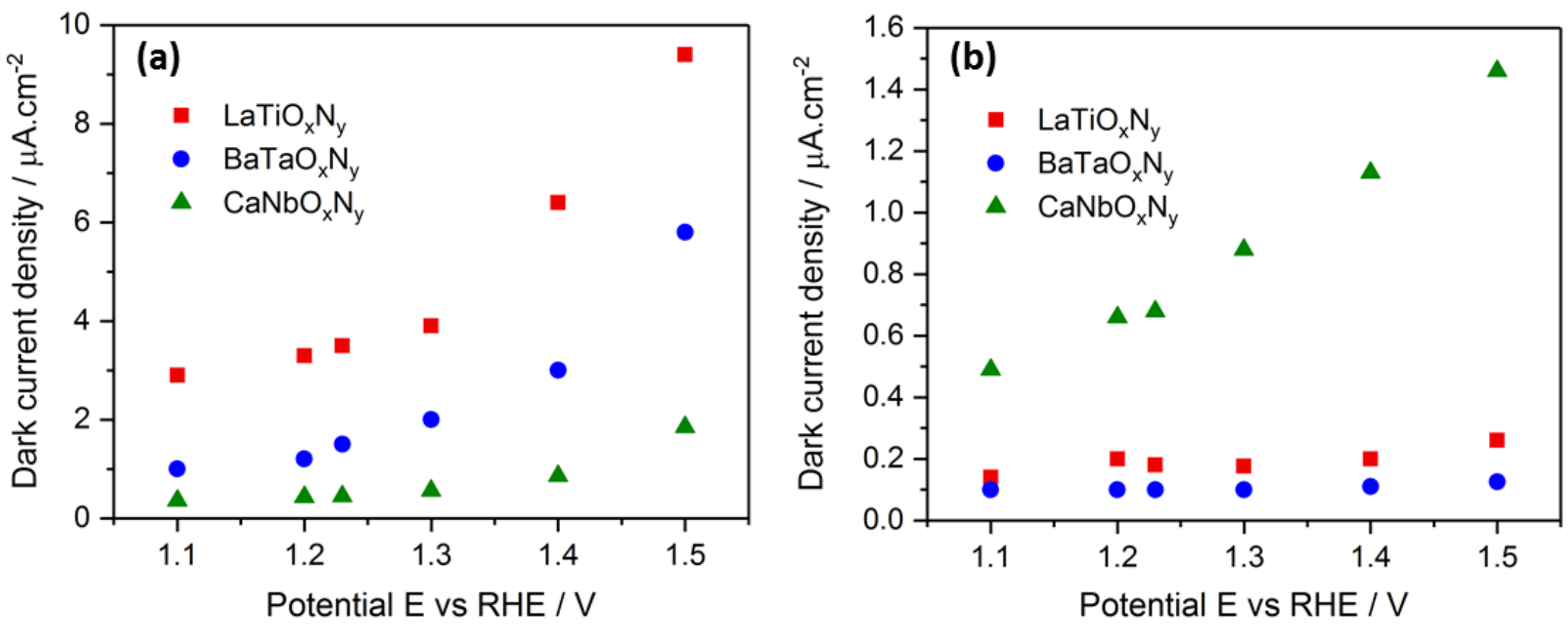

Figure S5. Dark current densities of $LaTiO_xN_y$, $BaTaO_xN_y$ and $CaNbO_xN_y$ (a) particle-based and (b) thin film photoanodes.

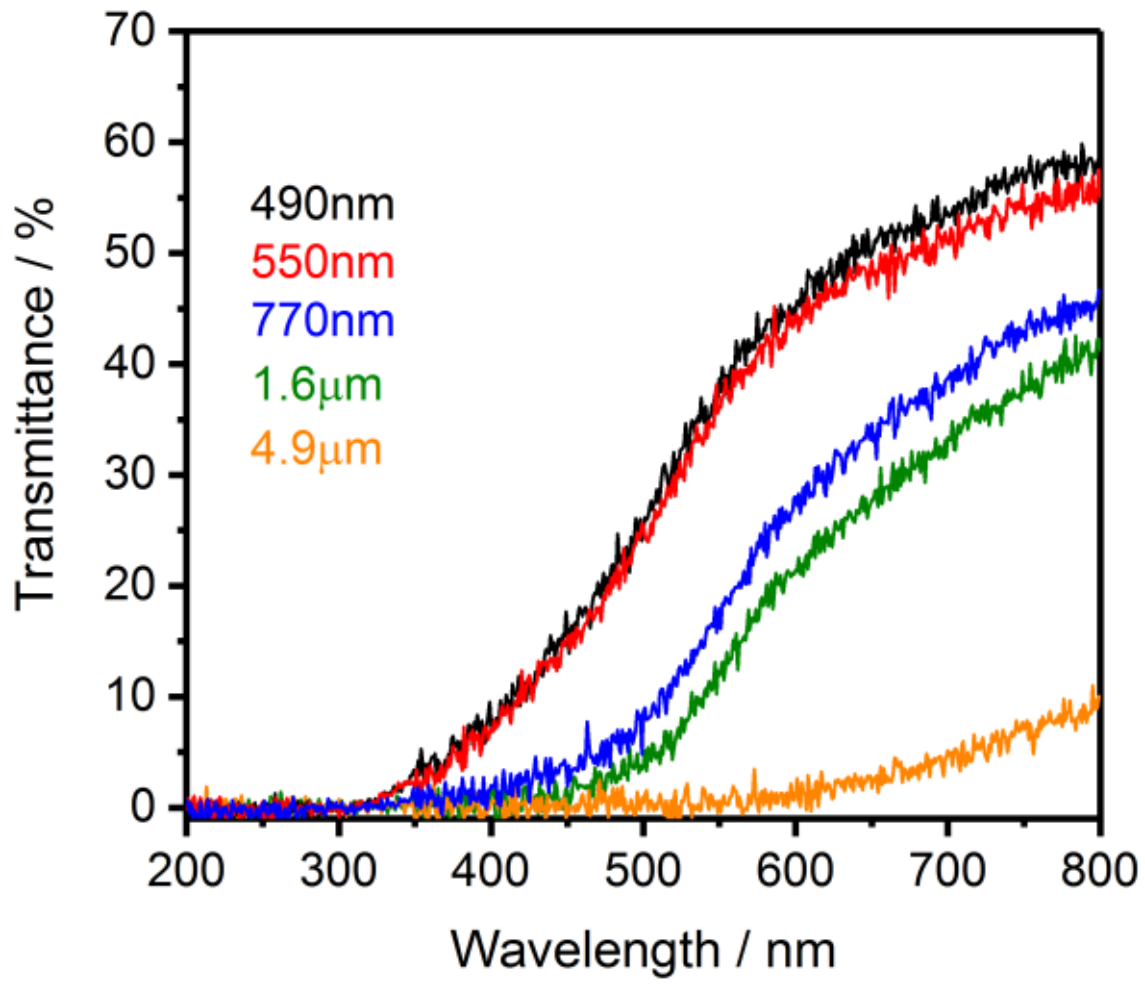

Figure S6. Transmittance spectra for $CaNbO_xN_y$ particulate photoanodes with different thicknesses.

Table S1. The photocurrents of the oxynitride thin films and particle-based photoanodes obtained from figure 5 corrected to the BET factors.

| Oxynitride | Photoanode | BET factor | J corrected to BET factor [μA.cm-2] |
|---|---|---|---|
| $LaTiO_xN_y$ | Thin Film | 1 | 0.6 |
| | Particle-based Photoanode | 16.15 | 0.59 |
| $BaTaO_xN_y$ | Thin Film | 1 | 0.633 |
| | Particle-based Photoanode | 12.75 | 0.43 |
| $CaNbO_xN_y$ | Thin Film | 1 | 0.139 |
| | Particle-based Photoanode | 22.1 | 0.138 |